\pdfoutput=1
\documentclass[aps,prd,reprint,nofootinbib,floatfix]{revtex4-2}

\usepackage{amsmath}
\usepackage{amssymb}
\usepackage{booktabs}
\usepackage{graphicx}
\usepackage{tikz}
\usetikzlibrary{shapes.geometric, arrows.meta, positioning}
\usepackage[colorlinks=true,linkcolor=blue,citecolor=blue,urlcolor=blue]{hyperref}
\usepackage{orcidlink}

\begin{document}

\title{Stratification Criteria for Machine Learning Pattern Discovery
       in Particle Physics: Preparing for the AlphaFold Moment}

\author{Andrew Michael Brilliant\,\orcidlink{0009-0004-8024-5442}}
\email{ab@ad-research.org}
\affiliation{Applied Dynamics Research, Sapporo, Japan}

\date{First posted 3 November 2025; this revision 15 June 2026}

\begin{abstract}
Machine learning capabilities are expanding into scientific domains at an
accelerating pace. When applied to high-energy physics pattern discovery,
they will generate candidates faster than traditional evaluation can absorb.

ML finds patterns in past data---it is inherently post hoc. Whether those
patterns reflect structure or coincidence is unknowable at discovery time;
this limitation applies equally to human and computational pattern-finding.
What differs is scale: ML candidate generation is effectively unbounded,
while human evaluation capacity remains fixed. When generation rate exceeds
evaluation bandwidth, binary accept/reject degenerates to random sampling.
Information-theoretically, the only response that preserves ranking under
finite evaluation budget is stratification. By focusing on stratification
rather than binary filtering, rule adjustments can be made retroactively,
thresholds tuned as results accumulate, and evaluation bandwidth focused on
top-ranked candidates.

This paper attempts to codify those criteria, proposing seven
computationally evaluable standards for stratifying ML-generated patterns.
The goal is not to deliver verdicts but to prioritize which candidates merit
pre-registration and longitudinal tracking. The framework preserves the
essential paradigm: pattern plus theory equals potentially real physics.
Patterns alone, however striking, remain candidates until theoretical
understanding arrives.

Making these criteria explicit enables prefiltering at scale while creating a
collaborative resource rather than a competitive one. ML capabilities extend
what physicists can search while preserving how physicists evaluate. We offer
this provisional framework for community calibration, with the goal of
developing validation infrastructure before the capability fully arrives.
\end{abstract}

% Keywords are deliberately not printed (the published version shows none).
% Paste these into the arXiv "Keywords" metadata field instead:
%   temporal convergence; pattern evaluation; lattice QCD;
%   machine learning; validation infrastructure

\maketitle

%=====================================================================
\section{Introduction}
\label{sec:intro}

Particle physics has developed sophisticated standards for evaluating
numerical patterns. When someone claims a ``striking'' relationship among
measured quantities, trained physicists know how to respond. They ask about
temporal priority: was this form predicted before seeing the data, or
discovered by searching? They probe for scale dependence: does the
relationship hold across energy scales, or only at one? They demand trial
accounting: how many functional forms were attempted before this one
succeeded? These questions reflect hard-won disciplinary knowledge about how
numerical coincidences arise and how to distinguish them from genuine
structure.

Portions of this evaluative expertise have remained tacit. It transmits
through apprenticeship, through referee reports, seminar questions, and
advisor feedback, rather than through explicit codification. High-energy
physics statistical practice already embodies Mayo's concept of ``severe
testing'' \cite{mayo2018,mayospanos2006} at an operational level, even when
practitioners do not consciously invoke the philosophical framework. HEP's
procedural apparatus---blind analyses, systematic uncertainty audits, Monte
Carlo coverage studies, and independent replication across
collaborations---constitutes severe testing in the error-statistical sense
(Section~\ref{sec:severity}). The standards exist; they have simply never
required formal articulation.

This paper is offered as a provisional framework for community calibration,
not a finished standard. The criteria proposed here represent one
researcher's attempt to make evaluation logic explicit; they are intended as
a starting point for discussion, not a prescriptive system. Several
thresholds are deliberately left coarse, and the sterile neutrino worked
example (Section~\ref{sec:neutrino}) would benefit from domain expert
calibration of the severity assessments. We invite corrections,
counterexamples, and alternative formulations. The value of this document
lies not in getting every criterion right on the first pass, but in making
the criteria explicit enough to be wrong in identifiable, correctable ways.

Machine learning changes this requirement. ML is increasingly used to search
high-dimensional parameter spaces in particle physics
\cite{feickert2021,karagiorgi2022}, generating pattern candidates at a rate
that can outpace human evaluation. Beyond Standard Model theories define
large parameter spaces in which ML systems can search at scale, producing
many candidate patterns---most of which will be spurious. The practitioners
deploying these systems may not have absorbed, through years of disciplinary
apprenticeship, the implicit standards that experienced physicists apply
automatically.

Why focus on mass relationships? The problem has a specific structure that
computational search is suited to exploit. The observables are few: nine
precisely measured fermion masses. The search space is vast: every algebraic
relationship among them, every transcendental expression, every
group-theoretic construction. And the literature is fragmented: Koide (1982)
\cite{koide1982}, Barut (1979) \cite{barut1979}, Foot (1994)
\cite{foot1994}, Rivero (2005) \cite{rivero2005}, Kocik (2012)
\cite{kocik2012}, each explored a small patch of this space by hand over
years. No individual researcher can survey the full combinatorial landscape.
This is the same structural bottleneck that made protein folding an ideal
target for computational methods \cite{jumper2021}: low-dimensional
observables, an enormous configuration space, and decades of partial human
exploration that collectively covered only a fraction of the possibilities.
Mass relationships are chosen here not because they are important (though
they are---the Standard Model accommodates masses but does not predict them)
but because the problem has the right shape for corpus-scale search to offer
a structural advantage over individual researchers.

Simple algebraic expressions relating measured masses represent precisely the
pattern type ML systems will generate at scale. They require no physical
insight to propose, only numerical search. The community response to
historical mass formulas illustrates the evaluation problem: genuine
interest, justified skepticism, and rarely resolution. This paper uses mass
relationships as a worked example not to advocate for any specific formula,
but because they cleanly illustrate evaluation challenges ML will amplify.

A second example provides something stronger than historical illustration.
The seven criteria were formalized before MicroBooNE \cite{microboone2025}
published its decisive test of the LSND sterile neutrino anomaly in late
2025, excluding the single light sterile neutrino interpretation of
LSND/MiniBooNE at 95\% CL.
The sterile neutrino anomalies---among the most prominent unresolved
questions in neutrino physics over the past two decades---therefore reached
experimental resolution after the framework existed, providing a rare case
where the criteria can be evaluated prospectively against known ground truth.

This paper attempts to define explicit criteria for this purpose. We propose
seven criteria for evaluating ML-generated pattern candidates, designed to
operationalize what trained physicists already know: that pre-data prediction
matters, that scale-dependent relations are suspect, that trial factors must
be reported, and that temporal convergence (stable or improving agreement as
measurement precision increases) functions as a severe test that coincidences
systematically fail.

Three features define the approach:

First, the framework addresses a specific decision: which patterns merit
pre-registration and longitudinal tracking across future data releases. The
output is not ``this pattern is correct'' but ``this pattern is worth
timestamping and waiting on.''

Second, the criteria are designed to be computationally evaluable where
possible, enabling application to large candidate sets. Continuous scores
rather than binary verdicts allow stratification across millions of
candidates.

Third, the framework explicitly acknowledges a precision regime problem.
Current quark mass determinations offer approximately 35,000 times less
discriminatory power than charged lepton masses
(Section~\ref{sec:discriminatory}). This gap means statistical agreement
alone, the traditional first filter, provides minimal evidence of structure
over coincidence. The look-elsewhere effect \cite{gross2010} that particle
physics takes seriously for mass peak searches applies equally to numerical
pattern searches; the trial factor problem is severe. Different filtering
criteria are needed, and those criteria must be explicit rather than
implicit.

The paper proceeds as follows. Section~\ref{sec:problem} characterizes the
pattern validation problem, including the discriminatory power gap that makes
quark phenomenology qualitatively different from lepton phenomenology.
Section~\ref{sec:precedent} examines a precedent from structural biology
where similar infrastructure questions arose.
Section~\ref{sec:framework} presents the seven criteria.
Section~\ref{sec:historical} illustrates the criteria against historical
cases. Section~\ref{sec:discussion} addresses implementation and limitations.

%=====================================================================
\section{The Pattern Validation Problem}
\label{sec:problem}

\subsection{Numerical Patterns in Physics}

The history of particle physics includes numerous proposed numerical
relationships among fundamental constants and particle masses. Some, like the
Gell-Mann-Okubo relation \cite{gellmann1961,okubo1962}, were eventually
explained by deeper theory. Others, such as Nambu's muon-electron mass ratio
\cite{nambu1952} and Lenz's proton-electron formula \cite{lenz1951}, appeared
compelling at contemporary precision but diverged as measurements improved.
Still others, including Koide's charged lepton mass formula \cite{koide1982}
and Barut's lepton mass relation \cite{barut1979}, have persisted for decades
without theoretical explanation.

This range of outcomes illustrates the core evaluation problem. Some numerical
patterns reflect genuine structure; others are coincidences exposed by
improving precision. The distinction is often unclear at the time of
discovery, and the community's evaluative standards for making this
distinction have remained largely implicit.

The distinction between ``real physics'' and ``mere formula'' is less clear
than it might seem. Amp\`ere and others developed descriptive equations for
electromagnetism decades before Maxwell unified them into a coherent theory.
Superconductivity was observed in 1911; the phenomenological equations proved
useful for engineering; the microscopic explanation (BCS theory) arrived in
1957. During those forty-six years, superconductivity was ``real physics''
despite lacking theoretical derivation. The patterns worked; the explanation
waited.

\subsection{The Discriminatory Power Gap}
\label{sec:discriminatory}

The precision available for testing patterns varies enormously across
observables. This variation has profound implications for how we should
evaluate claimed relationships.

\noindent\textbf{Charged Leptons:}

\begin{center}
\begin{tabular}{lcc}
\toprule
Particle & Mass & Precision \\
\midrule
Electron & 0.511 MeV   & $\sim 10^{-10}$ \\
Muon     & 105.7 MeV   & $\sim 10^{-8}$  \\
Tau      & 1776.9 MeV  & $\sim 10^{-4}$  \\
\bottomrule
\end{tabular}
\end{center}

\noindent The dynamic range spans $\tau/e \approx 3{,}500$. All masses are pole masses:
static, directly measurable, no scale dependence. Two anchors (electron,
muon) are known to extraordinary precision. With the worst precision being
the tau at 0.007\%, there are approximately \textbf{50 million
distinguishable positions} across this parameter space.

\noindent\textbf{Light Quarks:}

\begin{center}
\begin{tabular}{lcc}
\toprule
Particle & Mass (2 GeV) & Precision \\
\midrule
Up      & 2.16 MeV  & $\sim$3\%   \\
Down    & 4.7 MeV   & $\sim$1.5\% \\
Strange & 93.5 MeV  & $\sim$1\%   \\
\bottomrule
\end{tabular}
\end{center}

The dynamic range is only $s/u \approx 43$. All masses are
$\overline{\text{MS}}$ running masses. No anchors: all three have comparable
relative uncertainties. With the worst precision at 3\%, there are
approximately \textbf{1,400 distinguishable positions}.

\noindent\textbf{The ratio:}
$50{,}000{,}000 \,/\, 1{,}400 \approx 35{,}000\times$

This is an order-of-magnitude estimate, not a precision claim; the essential
point is robust. Leptons offer roughly $3.5 \times 10^{4}$ times greater
discriminatory power than light quarks.

% NOTE: the MDPI source repeated the "Historical intuitions calibrated on
% lepton phenomenology ... unremarkable among quarks" sentences twice --
% once here and again in the paragraph below. Kept once.
This asymmetry has direct implications for pattern evaluation. A pattern
achieving sub-percent agreement in the lepton sector survives a far more
stringent test than the same agreement in the quark sector. Historical
intuitions calibrated on lepton phenomenology systematically underestimate
coincidence rates in quark phenomenology. What would be striking among
leptons may be unremarkable among quarks.

This is why temporal convergence matters as a discriminator: when error bars
shrink across successive data releases, coincidences are exposed while
genuine relationships persist. We posit that temporal convergence functions
as a severe test in the sense formalized by Mayo \cite{mayo2018}: a
hypothesis gains credibility when it passes tests that would likely reveal
flaws if flaws existed.

Beyond precision, the lepton sector enjoys a structural advantage: charged
lepton masses run negligibly under QED ($\alpha \approx 1/137$), making pole
masses effectively scale-invariant. Quark masses run substantially under QCD
($\alpha_s \approx 0.1$--$0.3$), and relations that hold at one scale may
fail at another. This constraint is documented by Xing and Zhang
\cite{xing2006}. QCD running requires either specifying a scale or adding
correction terms, increasing the complexity of any proposed relationship. The
community's implicit caution toward quark mass formulas reflects sound
methodology: patterns with less precision, operating in a sector with less
discriminatory power, face a higher evidential bar.

\subsection{The Validation Bottleneck}

If genuine mass relationships exist among Standard Model observables,
improving precision will reveal them. Relationships that ``almost work'' at
2\% uncertainty will either crystallize into stable agreement or dissolve
into statistical noise as measurements improve.

This does not make ML-discovered patterns worthless; it means they require
validation infrastructure current practice has not developed.

Current peer review processes are designed for modest discovery rates. A
typical phenomenology paper might explore $10^{4}$ to $10^{6}$ parameter
points and report a handful of interesting configurations. If automated
systems generate thousands of pattern candidates, each with superficially
impressive numerical agreement, reviewers cannot evaluate claims faster than
they arrive.

The problem compounds when the same data informs both pattern discovery and
evaluation. This correlated error means evaluators tend to accept precisely
the spurious patterns the generator produces. Temporal convergence breaks
this correlation by using future, independent experimental determinations as
an external selection channel.

\subsection{The Theory Interface}

In high-energy physics terminology, machine learning operates in hep-ph
(phenomenology): it generates candidates. These candidates become physics
when they interface with hep-th (theory): when someone can explain why the
pattern holds.

This creates a bottleneck that databases alone cannot solve. Discovering
patterns is now easier than knowing what to do with them. Parameter space
search can produce dozens of sub-sigma candidates; introducing $\pi$ as a
permitted constant explodes the space further. Without stratification
criteria, resource allocation becomes guesswork.

The filtering criteria proposed here address this interface. The goal is to
identify which patterns merit pre-registration and longitudinal tracking:
candidates worth timestamping and waiting on, letting temporal convergence do
the discrimination that neither statistical fit nor theoretical intuition can
provide alone.

This addresses a specific historical moment. Lattice QCD has recently
achieved percent-level precision on hadron masses; each successive FLAG
review reports smaller uncertainties. If descriptive relationships exist in
quark physics, they might finally be findable. The ability to distinguish
genuine patterns from coincidences at scale is the prerequisite for answering
the question.

%=====================================================================
\section{A Precedent from Structural Biology}
\label{sec:precedent}

% database. AlphaFold2 (2021) demonstrated accurate prediction; the 200M+
% figure is the AlphaFold DB as of 2024. Split across the two citations.
In 2021, DeepMind's AlphaFold2 demonstrated highly accurate protein structure
prediction \cite{jumper2021}, and by 2024 the AlphaFold Protein Structure
Database contained over 200 million predicted structures \cite{varadi2024}.
The Protein Data Bank had accumulated approximately 180,000 experimentally
determined structures over fifty years. AlphaFold exceeded this by three
orders of magnitude in months.

% is the correct support for "frameworks for evaluating which predictions to
% trust".
This created an interface problem: ML systems generating candidates faster
than domain experts could evaluate them. The structural biology community
continues developing frameworks for evaluating which predictions to trust and
how to incorporate them into experimental workflows \cite{akdel2022}. These
validation standards emerged after the capability arrived.

AlphaFold had clear targets: experimentally determined 3D structures against
which predictions could be validated. Particle physics pattern discovery
lacks this clarity. A machine learning system might identify a relationship
among quark masses that achieves sub-sigma agreement with current
measurements. What does this mean? Is it genuine? A statistical artifact? A
coincidence destined to dissolve with improved precision?

High-energy physics has the opportunity to prepare such infrastructure in
advance. The framework proposed here is one attempt to do so.

%=====================================================================
\section{Framework: Seven Criteria}
\label{sec:framework}

\begin{figure}[tbp]
\centering
\resizebox{0.82\columnwidth}{!}{%
\begin{tikzpicture}[
    node distance=0.42cm and 2cm,
    box/.style={rectangle, draw, rounded corners, minimum width=3.5cm,
                minimum height=0.8cm, align=center, font=\small},
    gate/.style={diamond, draw, aspect=2.2, minimum width=1.5cm,
                 align=center, font=\scriptsize},
    fail/.style={rectangle, draw, fill=black!8, rounded corners,
                 minimum width=1.6cm, minimum height=0.6cm, align=center,
                 font=\scriptsize},
    arr/.style={-{Stealth[length=3pt]}, thick},
    passlbl/.style={font=\scriptsize, text=black!60},
    faillbl/.style={font=\scriptsize, text=black!60},
]
% Input
\node[box, fill=blue!8] (input) {ML Candidate\\Pattern};

% Criteria as gates
\node[gate, below=of input] (c1) {C1: Scale\\Invariance};
\node[gate, below=of c1] (c2) {C2:\\Compression};
\node[gate, below=of c2] (c3) {C3: Statistical\\Agreement};
\node[gate, below=of c3] (c4) {C4: Temporal\\Convergence};
\node[gate, below=of c4] (c5) {C5:\\Simplicity};
\node[gate, below=of c5] (c6) {C6: Independent\\Validation};
\node[gate, below=of c6] (c7) {C7: Theoretical\\Viability};

% Output
\node[box, below=of c7, fill=green!8] (output)
     {Stratified Ranking\\for Pre-registration};

% Fail nodes
\node[fail, right=of c1] (f1) {Filter};
\node[fail, right=of c2] (f2) {Filter};
\node[fail, right=of c3] (f3) {Filter};
\node[fail, right=of c4] (f4) {$T=-1$};
\node[fail, right=of c5] (f5) {Filter};
\node[fail, right=of c6] (f6) {Filter};
\node[fail, right=of c7] (f7) {Incompatible};

% Vertical arrows
\draw[arr] (input) -- (c1);
\draw[arr] (c1) -- node[passlbl, left] {pass} (c2);
\draw[arr] (c2) -- node[passlbl, left] {pass} (c3);
\draw[arr] (c3) -- node[passlbl, left] {pass} (c4);
\draw[arr] (c4) -- node[passlbl, left] {pass} (c5);
\draw[arr] (c5) -- node[passlbl, left] {pass} (c6);
\draw[arr] (c6) -- node[passlbl, left] {pass} (c7);
\draw[arr] (c7) -- node[passlbl, left] {pass} (output);

% Fail arrows
\draw[arr] (c1) -- node[faillbl, above] {fail} (f1);
\draw[arr] (c2) -- node[faillbl, above] {fail} (f2);
\draw[arr] (c3) -- node[faillbl, above] {fail} (f3);
\draw[arr] (c4) -- node[faillbl, above] {fail} (f4);
\draw[arr] (c5) -- node[faillbl, above] {fail} (f5);
\draw[arr] (c6) -- node[faillbl, above] {fail} (f6);
\draw[arr] (c7) -- node[faillbl, above] {fail} (f7);

\end{tikzpicture}}
\caption{Evaluation cascade for ML-generated pattern candidates. Each
criterion acts as a filter; candidates passing all seven receive stratified
ranking for pre-registration and longitudinal tracking. Criterion 4 (temporal
convergence) is the primary discriminator. The ordering is logical, not
strictly sequential: in practice, criteria may be evaluated in parallel where
data permits.}
\label{fig:workflow}
\end{figure}
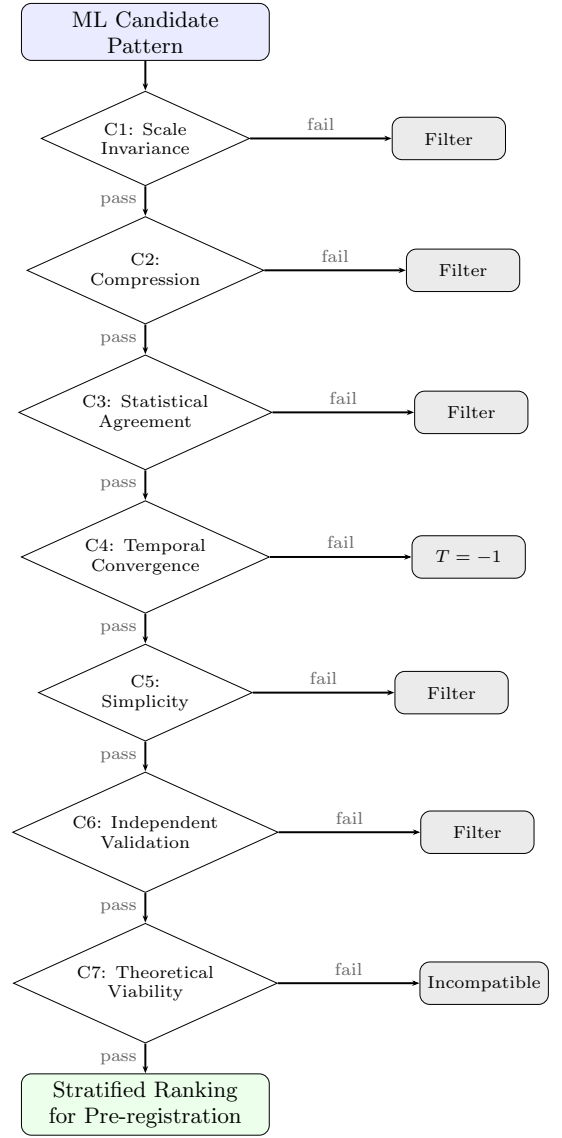

The specific thresholds proposed are initial estimates subject to community
calibration. Criteria 1 through 6 address empirical validation accessible to
phenomenologists. Criterion 7 addresses theoretical context.
Figure~\ref{fig:workflow} summarizes the evaluation cascade.

\subsection{Criterion 1: Scale Invariance Under Renormalization Group
            Evolution}

Mass ratios should ideally be scale-invariant under QCD running.
Scale-dependent relationships require justification: if a formula invokes a
specific energy scale, one must explain why that scale is privileged.
Otherwise, scale choice becomes a fine-tuning parameter.

\noindent\textbf{Empirical context:} Light quark mass ratios $m_s/m_d$ and
$m_d/m_u$ are preserved to high precision under QCD renormalization group
evolution. Relationships expressed as ratios automatically satisfy scale
invariance to the extent that running is flavor-universal at leading order.

\noindent\textbf{Proposed threshold:} Deviations $< 10^{-4}$ across 1 GeV to
TeV scales. Relationships not expressed as scale-invariant ratios require
explicit justification.

\subsection{Criterion 2: Compression of Degrees of Freedom}

Patterns must reduce $N$ parameters to fewer degrees of freedom through
unified constraints. A pattern that does not compress information provides no
predictive content.

\noindent\textbf{Empirical context:} The lepton mass formula compresses three
masses to two degrees of freedom: given any two masses, the third is
predicted. The Gell-Mann-Okubo relation similarly constrains hadron
multiplets. Compression is what distinguishes a pattern from a tautology.

\noindent\textbf{Proposed threshold:} $N$ parameters reduced to at most
$N-1$ degrees of freedom.

\subsection{Criterion 3: Statistical Agreement}

Patterns must agree with measurements within statistical bounds. This
criterion is necessary but not sufficient: given the low discriminatory power
in the quark sector, statistical agreement alone provides minimal evidence of
structure over coincidence.

\noindent\textbf{Proposed threshold:} $< 1\sigma$ deviation, with the
recognition that this threshold alone admits many coincidences at current
quark mass precision. Criterion 4 provides the primary filter.

\subsection{Criterion 4: Temporal Convergence}

Patterns must demonstrate directional convergence or stability across data
releases. This is the core discriminator, operationalizing Mayo's severity
criterion \cite{mayo2018,mayospanos2011}: a claim gains credibility only by
passing tests that would probably have revealed flaws if flaws existed.
Severity is a property of the \textit{testing procedure}, not merely of the
numerical outcome, a distinction developed in Section~\ref{sec:severity}.
Improving experimental precision constitutes a severe test because it has
high probability of exposing spurious agreement: coincidences systematically
fail it. When error bars shrink, genuine patterns show residuals shrinking or
stable. Coincidences show residuals growing: precision exposes the latent
offset that broader uncertainties masked.

\noindent\textbf{Requirements:}
\begin{itemize}
\item Pre-registration via timestamped repository before new data releases
\item Central values converging toward (or stable around) prediction as
      precision improves
\end{itemize}

\noindent\textbf{Why temporal convergence provides robust protection:}

Pre-registration creates an immutable record. Authors cannot select data
vintages post-hoc, adjust formulas after seeing results, or claim prescience
after observing convergence. The only way to pass Criterion 4 is genuine
predictive success across independent experimental cycles.

\noindent\textbf{Empirical context:} The lepton mass relationship has
survived four decades of improving measurements. Historical patterns that
diverged (Section~\ref{sec:historical}) demonstrate the converse: initial
statistical agreement revealed as coincidental when precision improved.

\noindent\textbf{Proposed threshold:} Convergence or stability demonstrated
across $\geq 3$ independent data releases, with pre-registration established
before each release. Patterns showing systematic divergence receive $T = -1$
classification and are filtered from further consideration.

\subsection{Criterion 5: Mathematical Simplicity}

Complexity overfits. Simplicity constrains hypothesis space.

\noindent\textbf{Empirical context:} Unconstrained complexity fits any
relationship among any quantities: it explains everything and predicts
nothing.

\noindent\textbf{Proposed threshold:} Basic arithmetic operations, integer
exponents $\leq 5$, standard constants ($\pi$, small integers). Formulas
requiring evaluation at a specific renormalization scale that is not
derivable from the relationship itself incur an additional degree of freedom
in the complexity count; a scale-invariant formula is inherently simpler than
one requiring a specified $\mu$. This threshold requires sharper
operationalization; the principle is sound even if the boundary needs
calibration.

\subsection{Criterion 6: Independent Validation}

Patterns must show consistency across independent determinations. A pattern
appearing in one collaboration's results but not others may reflect
systematic artifacts.

\noindent\textbf{Empirical context:} FLAG 2024 \cite{flag2024} averages draw
on results from BMW, MILC, HPQCD, ETM, RBC/UKQCD, and other collaborations
employing different lattice discretizations, fermion actions, and analysis
methods.

\noindent\textbf{Proposed threshold:} Agreement across $\geq 3$ independent
collaborations or methods with demonstrably different systematic
uncertainties.

\subsection{Criterion 7: Theoretical Viability}

Patterns must not be demonstrably incompatible with established physics. This
criterion requires theoretical rather than empirical assessment.

\noindent\textbf{Three possible outcomes:}
\begin{itemize}
\item \textbf{PASS (Compatible):} Mechanism identified within existing
      frameworks.
\item \textbf{PASS (Unknown):} No known incompatibility. Pattern awaits
      theoretical investigation.
\item \textbf{FAIL (Incompatible):} Pattern contradicts established
      constraints through explicit proof.
\end{itemize}

\noindent\textbf{Critical:} Absence of mechanism does not constitute failure.
The lepton mass formula remains in Unknown status despite decades of
attention. Multiple theoretical mechanisms have been proposed; none have
achieved consensus. This has not diminished the formula's empirical standing.

%=====================================================================
\section{Historical Illustration}
\label{sec:historical}

To demonstrate how the criteria operate, we apply them to historical cases.
We emphasize that alignment between criteria and historical outcomes is
expected by construction: criteria were partly informed by examining patterns
that persisted. This provides illustration, not validation.

A contemporary example offers something stronger. The sterile neutrino
anomalies reached experimental resolution after the framework was formalized
(Section~\ref{sec:neutrino}). Unlike the historical cases above, this
constitutes a prospective test: the criteria existed before the ground truth
arrived, and were not adjusted after examining the outcome.

\subsection{Patterns That Converged}

\noindent\textbf{Gell-Mann-Okubo Relation} \cite{gellmann1961,okubo1962}:
Predicted hadron mass relationships before the quark model existed.

\begin{center}
\small
\begin{tabular}{ll}
\toprule
Criterion & Assessment \\
\midrule
1. Scale Invariance & PASS \\
2. Compression      & PASS \\
3. Statistical      & PASS \\
4. Temporal         & PASS: validated by subsequent data \\
5. Simplicity       & PASS \\
6. Independent      & PASS \\
7. Theoretical      & Unknown $\rightarrow$ Explained (SU(3)) \\
\bottomrule
\end{tabular}
\end{center}

GMO demonstrates that patterns can legitimately persist in Unknown status
until theory catches up. The empirical pattern preceded the theoretical
explanation by years.

\noindent\textbf{Lepton Mass Formula} \cite{koide1982}: Passes all empirical
criteria, remains in Unknown theoretical status after four decades.

\subsection{Patterns That Diverged}

\noindent\textbf{Nambu (1952)} \cite{nambu1952}:
$m_\mu/m_e = 3/(2\alpha)$ (modern rendering of Nambu's empirical mass
spectrum)

Initially $\sim 1\sigma$ agreement. Now $> 20\sigma$ deviation. Temporal
convergence: $T = -1$, diverged.

\noindent\textbf{Lenz (1951)} \cite{lenz1951}: $m_p/m_e \approx 6\pi^{5}$

Initially $< 1\sigma$ agreement. Now $\sim 50\sigma$ deviation. Temporal
convergence: $T = -1$, diverged.

These were well-motivated given contemporary precision. They were superseded
by improved measurement, not refuted by argument. The temporal convergence
criterion correctly identifies both as diverging.

\subsubsection{Sterile Neutrinos (Worked Example)}
\label{sec:neutrino}

The sterile neutrino anomalies offer a qualitatively different worked
example: an experimental anomaly (excess events interpreted as evidence for
new physics) that reached definitive resolution in 2025, after the framework
criteria were formalized. This temporal sequence is critical: the criteria
were not designed to accommodate this outcome, and were not adjusted after
reviewing it. The neutrino case therefore serves as a prospective rather than
retrospective test of the framework.

An important clarification is necessary. The physics community resolved the
sterile neutrino anomaly through its own established practices: independent
replication, improved detector technology, global fits across channels, and
sustained skepticism toward claims that could not self-consistently
accommodate all available data. The framework proposed here attempts to
codify that existing practice into explicit, computationally evaluable
criteria. This worked example illustrates that the community's implicit
procedures align with the criteria we document, not the reverse. The purpose
is not to evaluate past practice but to replicate it outside traditional
channels, at the scale that ML-generated candidates will require.

% 4.7 sigma with antineutrino data. The 6.0 sigma is LSND + MiniBooNE.
% Also: LSND measured an excess; the sterile interpretation came after.
\paragraph{The anomaly.} In 2001, LSND reported a $3.8\sigma$ excess
consistent with $\bar{\nu}_\mu \to \bar{\nu}_e$ oscillations with
$\Delta m^2 \sim 0.2$--$10$~eV$^2$ \cite{lsnd2001}, subsequently interpreted
within the sterile neutrino framework. In 2018, MiniBooNE reported a
$4.5\sigma$ $\nu_e$ appearance excess in neutrino mode---rising to
$4.7\sigma$ when combined with antineutrino data---and the combined
LSND~+~MiniBooNE excess reached $6.0\sigma$ \cite{miniboone2018}. Two
experiments, different detector technologies, apparently converging on the
same signal.

\paragraph{Application of criteria.} At the point of maximum apparent
significance (2018--2020), how would the framework have assessed the sterile
neutrino claim?

% Wide (three prose columns) -- spans both columns in the two-column layout.
\begin{table*}[t]
\centering
\small
\caption{Framework assessment of the sterile neutrino claim at the point of
maximum apparent significance (c.\ 2020).}
\label{tab:sterile}
\begin{tabular}{@{}lll@{}}
\toprule
Criterion & Assessment (c.\ 2020) & Notes \\
\midrule
1. Scale Inv.  & Ambiguous     & Signal at specific $L/E$; untested across scales \\
2. Compression & PASS          & 2 parameters $\to$ 4 anomalies \\
3. Statistical & PASS          & $6.0\sigma$ combined \\
4. Temporal    & Ambiguous     & See below \\
5. Simplicity  & PASS          & Minimal extension (3+1 model) \\
6. Independent & \textbf{FAIL} & ICARUS null \cite{icarus2013};
                                 appearance--disappearance tension \\
7. Theoretical & \textbf{STRESSED} & Global fits showed internal tension
                                 $> 4\sigma$ \cite{diaz2020} \\
\bottomrule
\end{tabular}
\end{table*}

The critical diagnostics were Criteria~6 and~7. While LSND and MiniBooNE
appeared to confirm each other, ICARUS found results inconsistent with LSND's
preferred parameters \cite{icarus2013}, and global fits incorporating both
appearance and disappearance channels showed persistent internal tension: the
mixing parameters required to explain the appearance signal predicted
disappearance effects that were not observed \cite{diaz2020}.

Criterion~4 appeared superficially favorable (two experiments over
17~years), but the two experiments, while at different laboratories (Los
Alamos and Fermilab) and different beam energies, shared similar $L/E$
baselines and analogous detector limitations (both relied on Cherenkov-based
detection unable to distinguish electrons from single photons). These
correlated detection systematics meant that MiniBooNE did not constitute a
genuinely independent temporal test in the sense the criterion requires.

% stated explicitly rather than left as a bare Delta-chi-squared.
\paragraph{Resolution.} In late 2025, MicroBooNE published a decisive test of
the LSND anomaly using liquid argon TPC technology, capable of resolving the
electron/photon ambiguity that limited both LSND and MiniBooNE, and
simultaneously analyzed two neutrino beams to break
appearance--disappearance degeneracies \cite{microboone2025}. Result:
$\Delta\chi^2 = 0.228$ relative to the three-neutrino model, excluding the
single light sterile neutrino interpretation of LSND/MiniBooNE at 95\% CL. No
evidence for sterile neutrinos in the $\Delta m^2 \sim 1$~eV$^2$ parameter
space relevant to LSND.\footnote{We note that KATRIN \cite{katrin2025} independently
constrains sterile neutrino mixing through tritium $\beta$-decay kinematics
in a different region of parameter space
($\Delta m^2 \gtrsim 10$~eV$^2$). While part of the broader sterile neutrino
program, KATRIN addresses a distinct anomaly from LSND.}

\paragraph{What the framework captures.} The sterile neutrino hypothesis
\textit{passed} criteria requiring only a plausible model and a significant
signal (Criteria~2, 3, 5). It \textit{failed} criteria demanding consistency
across genuinely independent tests (Criterion~6) and self-consistent
theoretical interpretation (Criterion~7). The appearance--disappearance
tension visible in global fits by 2020 was the framework's equivalent of a
diverging residual.

The sharpest lesson concerns Criterion~4. The LSND$\to$MiniBooNE sequence
\textit{appeared} to show convergence, but the two experiments shared
analogous detector technology (both relied on Cherenkov detection unable to
resolve the electron/photon ambiguity), meaning that MiniBooNE did not
constitute a genuinely independent test despite its statistical significance.
Genuine temporal convergence requires that each successive test be capable of
falsifying the claim. MicroBooNE's liquid argon TPC technology provided the
genuinely independent test; the claim did not survive it. Temporal
convergence: $T = -1$.

\paragraph{Methodological note.} Applying the criteria to the state of
evidence circa 2020 is retrospective. However, the experimental resolution
arrived after the framework was formalized, making the outcome a prospective
test: the criteria that would have flagged Criteria~6 and~7 as stressed were
not constructed to produce this result. We make no claim that the framework
would have prevented the anomaly from receiving attention, nor should it
have. The sterile neutrino case serves a different purpose: it tests whether
the framework accurately describes how the HEP community already evaluates
anomalies. The criteria that flagged the LSND signal as evidentially stressed
(Criteria~6 and~7) correspond to concerns that experimentalists and theorists
were already raising informally by 2020. The framework did not discover these
concerns; it codified them. That the codification produces the same
assessment as expert judgment, prospectively confirmed by MicroBooNE's
resolution, supports the premise that implicit evaluation standards can be
made explicit without distortion. The goal is then to apply this codified
description to contexts where ML-generated candidates arrive faster than
expert evaluation can absorb them, not to prescribe how domain experts should
evaluate evidence, but to document their practice in a form that scales.

\subsection{The Value of Historical Analysis}

In this limited sample: all patterns that persisted show $T \geq 0$; all that
diverged show $T = -1$. This alignment is expected, as criteria were informed
by these outcomes. The test of the framework is prospective: will future
patterns classified as $T = +1$ persist, and will those classified as
$T = -1$ diverge?

We explicitly decline to ``validate'' this framework against the historical
mass-formula cases. Such testing would be post-hoc, cross-regime (leptons
offer 35,000$\times$ greater discriminatory power), and survivorship-biased.
Historical cases motivate the problem; they cannot validate whether explicit
criteria improve upon implicit evaluation. The sterile neutrino case is the
one partial exception: the experimental resolution arrived after the
framework was formalized, providing a prospective rather than retrospective
test.

%=====================================================================
\section{Discussion}
\label{sec:discussion}

\subsection{Theoretical Foundations}

The temporal convergence criterion implements external selection
\cite{brilliant2026esp} to break correlated error between pattern generation
and evaluation. When the same data informs both discovery and validation,
coincidental agreements persist; independent future data provides the
separation needed for meaningful tests.

The underlying principle---that genuine results survive repeated, independent
testing while spurious ones do not---is not new. It is arguably the central
insight of the philosophy of science. Popper's falsificationism holds that
scientific claims gain corroboration precisely by surviving tests that could
have refuted them \cite{popper1959}. Lakatos sharpened this into the
distinction between progressive research programmes, where ``theory leads to
the discovery of hitherto unknown novel facts,'' and degenerating ones that
merely accommodate existing data \cite{lakatos1970}. Mayo's severe testing
framework provides the modern statistical formalization: claims gain
credibility only by ``passing tests that probably would have found flaws,
were they present'' \cite{mayo2018}. The concept that convergence across
independent experiments constitutes strong evidence is, in this sense,
standard scientific methodology.

What temporal convergence adds is not the principle but its codification as a
computationally evaluable criterion for a specific problem: filtering
ML-generated pattern candidates at scale. When candidate volume is bounded by
human intuition, scientists apply these principles implicitly through peer
review, replication, and professional judgment. When ML generation produces
candidates faster than human evaluation can absorb, the implicit standard
must become explicit and automatable. The $T \in \{-1, 0, +1\}$ scoring, the
pre-registration protocol tied to specific data releases (e.g., FLAG
reviews), and the integration with the other six criteria constitute the
operational contribution---not the philosophical insight that surviving
independent tests matters.

A related statistical approach is Efron's false discovery rate (FDR)
framework \cite{efron2010}, which controls the expected proportion of false
positives among rejected hypotheses when testing many candidates
simultaneously. FDR addresses multiple testing within a single dataset;
temporal convergence addresses a complementary problem: longitudinal
evaluation across independent datasets released over time. A pattern may
survive FDR correction at a single time point yet diverge under subsequent
precision improvements, or conversely, a marginal candidate at one time point
may show progressive convergence as data accumulate. The two approaches are
complementary, not competing.

The pre-registration movement in psychology \cite{osc2015,nosek2018} provides
institutional precedent for such infrastructure.

\subsection{Error-Statistical Foundations and the Severity Principle}
\label{sec:severity}

The framework's reliance on severity warrants brief philosophical grounding.
Mayo and Spanos \cite{mayospanos2006,mayospanos2011} developed the
error-statistical approach as a meta-methodological principle: a hypothesis
$H$ passes a severe test $T$ with data $x$ only if (i)~$x$ agrees with $H$,
and (ii)~the test procedure had a high probability of producing a result that
\textit{does not agree} with $H$ if $H$ is false. Severity is a property of
the testing \textit{procedure}, not of the numerical output alone
\cite{cousins2020,fletcher2020}. For the purposes of this framework, we
require only the claim that improving experimental precision constitutes a
severe test of numerical patterns, because coincidences have high probability
of being exposed as measurement uncertainty shrinks.

Criterion~4 (Temporal Convergence) operationalizes this principle.
Pre-registration before independent data releases ensures that the testing
procedure has genuine error-probing capacity: a spurious pattern cannot
adjust to new data, and improving precision will expose the latent offset
that broader uncertainties masked. The sterile neutrino case
(Section~\ref{sec:neutrino}) illustrates the failure mode: MiniBooNE's
apparent confirmation of LSND was not a severe test in the procedural sense,
because analogous detector technology meant the procedure lacked the capacity
to discriminate the claimed signal from backgrounds. MicroBooNE's liquid
argon technology restored severity by providing genuinely independent
error-probing capacity.

\subsection{Practical Implementation}

Existing infrastructure can support pattern documentation without requiring
new systems. Zenodo, operated by CERN, provides timestamped DOI assignment
for predictions. A researcher identifying a candidate pattern can deposit the
formulation before the next FLAG release; the DOI provides immutable
pre-registration. Subsequent evaluation against new data constitutes the
temporal convergence test.

The lattice QCD community's FLAG collaboration demonstrates how distributed
expertise can produce authoritative consensus; similar structures might
emerge for pattern evaluation if the need becomes sufficient.

The goal is augmentation, not replacement. Peer review remains essential for
theoretical evaluation, methodological scrutiny, and contextual judgment. The
criteria proposed here add upstream structure: a pre-filter that reduces the
volume reaching human reviewers while increasing the proportion warranting
serious attention.

\subsection{Limitations}

\noindent\textbf{Temporal data requirements:} Criterion 4 requires multiple
data releases. New patterns cannot be fully evaluated immediately.

\noindent\textbf{Threshold calibration:} All proposed thresholds are initial
estimates. Community input determines appropriate values.

\noindent\textbf{Theoretical coupling subjectivity:} Criterion 7 requires
domain expertise and involves judgment.

\noindent\textbf{Domain specificity:} The criteria proposed here are
calibrated to particle physics phenomenology, specifically mass relationships
testable against lattice QCD and precision measurement data. Adaptation to
other domains (collider event classification, dark matter search strategies,
cosmological parameter estimation) would require separate development. We do
not claim generality beyond the specific use case addressed.

These are not flaws but acknowledgments. We propose a provisional framework,
not a finished system. We invite the community to treat this framework as we
propose treating empirical patterns: provisionally, with temporal tracking.

%=====================================================================
\section{Conclusion}

Machine learning will generate particle physics pattern candidates faster
than traditional evaluation can absorb them. The structural biology precedent
suggests that validation infrastructure is better developed before this
capability fully arrives than after.

This framework proposes explicit, computationally evaluable criteria for a
specific decision: which patterns merit pre-registration and longitudinal
tracking. The output is not ``ready for peer review'' but ``worth
timestamping and waiting,'' letting temporal convergence distinguish genuine
structure from coincidence across future data releases.

The goal is augmentation. Peer review remains essential; these criteria add
upstream structure that provides a basis for prioritization. The lattice QCD
community's decades of precision work are not threatened by pattern-discovery
approaches; they are essential to them. Without reliable ground truth,
patterns cannot be validated.

This is a starting point, not a solution. If the community finds specific
criteria too strict or certain thresholds irrelevant, that discussion is
progress. We offer these criteria as a basis for community calibration.

%=====================================================================
\begin{acknowledgments}
The author thanks Robert D.~Cousins for valuable feedback on an earlier
draft, and Riccardo M.~Pagliarella for encouragement and valuable
discussions. This work would not be possible without the extraordinary
precision achieved by the lattice QCD community, particularly FLAG, BMW,
MILC, HPQCD, and ETM collaborations.
\end{acknowledgments}

%=====================================================================
\section*{Declarations}

\noindent\textbf{Funding:} This research was conducted independently without
institutional funding or external support.

\noindent\textbf{Competing interests:} The author declares no competing
interests.

\noindent\textbf{Data availability:} All numeric values used in this analysis
are derived from publicly available FLAG 2024 \cite{flag2024} and PDG 2024
\cite{pdg2024} reviews.

\noindent\textbf{Code availability:} Analysis methodologies are documented at
\url{https://github.com/AndBrilliant/TemporalConvergence}.

\noindent\textbf{Author contributions:} A.M.B. conceived the framework,
performed the analysis, and wrote the manuscript.

\noindent\textbf{Publication note:} A version of this work was published in
the \textit{London Journal of Physics} \textbf{3}(1) (2026),
DOI 10.69710/ljp.v3i1.17825, under a Creative Commons CC BY 4.0 license; the
author retains full copyright. The present version has been revised for
presentation and citation accuracy. No criterion, threshold, or conclusion
differs from the published text.

%=====================================================================


%apsrev4-2.bst 2019-01-14 (MD) hand-edited version of apsrev4-1.bst
%Control: key (0)
%Control: author (8) initials jnrlst
%Control: editor formatted (1) identically to author
%Control: production of article title (0) allowed
%Control: page (0) single
%Control: year (1) truncated
%Control: production of eprint (0) enabled
\begin{thebibliography}{0}%
\makeatletter
\providecommand \@ifxundefined [1]{%
 \@ifx{#1\undefined}
}%
\providecommand \@ifnum [1]{%
 \ifnum #1\expandafter \@firstoftwo
 \else \expandafter \@secondoftwo
 \fi
}%
\providecommand \@ifx [1]{%
 \ifx #1\expandafter \@firstoftwo
 \else \expandafter \@secondoftwo
 \fi
}%
\providecommand \natexlab [1]{#1}%
\providecommand \enquote  [1]{``#1''}%
\providecommand \bibnamefont  [1]{#1}%
\providecommand \bibfnamefont [1]{#1}%
\providecommand \citenamefont [1]{#1}%
\providecommand \href@noop [0]{\@secondoftwo}%
\providecommand \href [0]{\begingroup \@sanitize@url \@href}%
\providecommand \@href[1]{\@@startlink{#1}\@@href}%
\providecommand \@@href[1]{\endgroup#1\@@endlink}%
\providecommand \@sanitize@url [0]{\catcode `\\12\catcode `\$12\catcode
  `\&12\catcode `\#12\catcode `\^12\catcode `\_12\catcode `\%12\relax}%
\providecommand \@@startlink[1]{}%
\providecommand \@@endlink[0]{}%
\providecommand \url  [0]{\begingroup\@sanitize@url \@url }%
\providecommand \@url [1]{\endgroup\@href {#1}{\urlprefix }}%
\providecommand \urlprefix  [0]{URL }%
\providecommand \Eprint [0]{\href }%
\providecommand \doibase [0]{https://doi.org/}%
\providecommand \selectlanguage [0]{\@gobble}%
\providecommand \bibinfo  [0]{\@secondoftwo}%
\providecommand \bibfield  [0]{\@secondoftwo}%
\providecommand \translation [1]{[#1]}%
\providecommand \BibitemOpen [0]{}%
\providecommand \bibitemStop [0]{}%
\providecommand \bibitemNoStop [0]{.\EOS\space}%
\providecommand \EOS [0]{\spacefactor3000\relax}%
\providecommand \BibitemShut  [1]{\csname bibitem#1\endcsname}%
\let\auto@bib@innerbib\@empty
%</preamble>
\end{thebibliography}%


\begin{thebibliography}{99}

\bibitem{microboone2025}
MicroBooNE Collaboration,
``Search for light sterile neutrinos with two neutrino beams at MicroBooNE,''
Nature \textbf{648}, 64--69 (2025), DOI: 10.1038/s41586-025-09757-7.

\bibitem{katrin2025}
KATRIN Collaboration,
``Sterile-neutrino search based on 259 days of KATRIN data,''
Nature \textbf{648}, 70--75 (2025), DOI: 10.1038/s41586-025-09739-9.

\bibitem{feickert2021}
M.~Feickert and B.~Nachman,
``A Living Review of Machine Learning for Particle Physics,''
arXiv:2102.02770 [hep-ph] (2021, continuously updated).

\bibitem{karagiorgi2022}
G.~Karagiorgi, G.~Kasieczka, S.~Kravitz, B.~Nachman, and D.~Shih,
``Machine learning in the search for new fundamental physics,''
Nat.\ Rev.\ Phys.\ \textbf{4}, 399--412 (2022),
arXiv:2112.03769 [hep-ph], DOI: 10.1038/s42254-022-00455-1.

\bibitem{gross2010}
E.~Gross and O.~Vitells,
``Trial factors for the look elsewhere effect in high energy physics,''
Eur.\ Phys.\ J.\ C \textbf{70}, 525--530 (2010),
arXiv:1005.1891 [physics.data-an], DOI: 10.1140/epjc/s10052-010-1470-8.

\bibitem{jumper2021}
J.~Jumper \textit{et al.},
``Highly accurate protein structure prediction with AlphaFold,''
Nature \textbf{596}, 583--589 (2021),
DOI: 10.1038/s41586-021-03819-2.

\bibitem{varadi2024}
M.~Varadi \textit{et al.},
``AlphaFold Protein Structure Database in 2024: providing structure coverage
for over 214 million protein sequences,''
Nucleic Acids Res.\ \textbf{52}, D368--D375 (2024),
DOI: 10.1093/nar/gkad1011.

\bibitem{akdel2022}
M.~Akdel \textit{et al.},
``A structural biology community assessment of AlphaFold2 applications,''
Nat.\ Struct.\ Mol.\ Biol.\ \textbf{29}, 1056--1067 (2022),
DOI: 10.1038/s41594-022-00849-w.

% published text's "84, 1263 (2024)" points at nothing.
\bibitem{flag2024}
Y.~Aoki \textit{et al.} (Flavour Lattice Averaging Group, FLAG),
``FLAG Review 2024,''
Eur.\ Phys.\ J.\ C \textbf{85}, 657 (2025),
arXiv:2411.04268 [hep-lat], DOI: 10.1140/epjc/s10052-025-14172-3.

\bibitem{pdg2024}
S.~Navas \textit{et al.} (Particle Data Group),
``Review of Particle Physics,''
Phys.\ Rev.\ D \textbf{110}, 030001 (2024),
DOI: 10.1103/PhysRevD.110.030001.

% Nuovo Cim. reference. Both papers are now cited, correctly paired.
\bibitem{koide1982}
Y.~Koide,
``Fermion-Boson Two-Body Model of Quarks and Leptons and Cabibbo Mixing,''
Lett.\ Nuovo Cim.\ \textbf{34}, 201--205 (1982),
DOI: 10.1007/BF02817096;
see also Y.~Koide, ``A New View of Quark and Lepton Mass Hierarchy,''
Phys.\ Rev.\ D \textbf{28}, 252 (1983),
DOI: 10.1103/PhysRevD.28.252.

\bibitem{xing2006}
Z.-Z.~Xing and H.~Zhang,
``On the Koide-like relations for the running masses of charged leptons, neutrinos and quarks,''
Phys.\ Lett.\ B \textbf{635}, 107--111 (2006),
arXiv:hep-ph/0602134, DOI: 10.1016/j.physletb.2006.02.060.

% the heavy-quark Q value cited against it in Sec. 2.2.
\bibitem{gellmann1961}
M.~Gell-Mann,
``The Eightfold Way: A Theory of Strong Interaction Symmetry,''
Caltech Synchrotron Laboratory Report CTSL-20 (1961);
reprinted in M.~Gell-Mann and Y.~Ne'eman, \textit{The Eightfold Way}, W.~A.~Benjamin, New York (1964);
see also M.~Gell-Mann, ``Symmetries of Baryons and Mesons,'' Phys.\ Rev.\ \textbf{125}, 1067--1084 (1962),
DOI: 10.1103/PhysRev.125.1067.

\bibitem{okubo1962}
S.~Okubo,
``Note on Unitary Symmetry in Strong Interactions,''
Prog.\ Theor.\ Phys.\ \textbf{27}, 949--966 (1962),
DOI: 10.1143/PTP.27.949.

\bibitem{nambu1952}
Y.~Nambu,
``An Empirical Mass Spectrum of Elementary Particles,''
Prog.\ Theor.\ Phys.\ \textbf{7}, 595--596 (1952),
DOI: 10.1143/PTP.7.5.595.

\bibitem{lenz1951}
F.~Lenz,
``The Ratio of Proton and Electron Masses,''
Phys.\ Rev.\ \textbf{82}, 554 (1951),
DOI: 10.1103/PhysRev.82.554.2.

\bibitem{barut1979}
A.~O.~Barut,
``Lepton Mass Formula,''
Phys.\ Rev.\ Lett.\ \textbf{42}, 1251 (1979),
DOI: 10.1103/PhysRevLett.42.1251.

\bibitem{rivero2005}
A.~Rivero and A.~Gsponer,
``The strange formula of Dr.\ Koide,''
arXiv:hep-ph/0505220 (2005).

\bibitem{foot1994}
R.~Foot,
``A note on Koide's lepton mass relation,''
arXiv:hep-ph/9402242 (1994), McGill/94-9.

\bibitem{kocik2012}
J.~Kocik,
``The Koide Lepton Mass Formula and Geometry of Circles,''
arXiv:1201.2067 [physics.gen-ph] (2012).

\bibitem{lakatos1970}
I.~Lakatos,
``Falsification and the Methodology of Scientific Research Programmes,''
in \textit{Criticism and the Growth of Knowledge}, eds.\ I.~Lakatos and A.~Musgrave,
Cambridge University Press, pp.\ 91--196 (1970),
DOI: 10.1017/CBO9781139171434.009.

\bibitem{popper1959}
K.~R.~Popper,
\textit{The Logic of Scientific Discovery},
Hutchinson \& Co., London (1959); Routledge Classics reprint (2002),
DOI: 10.4324/9780203994627, ISBN 978-0-415-27844-7.

\bibitem{efron2010}
B.~Efron,
\textit{Large-Scale Inference: Empirical Bayes Methods for Estimation, Testing, and Prediction},
Institute of Mathematical Statistics Monographs, Cambridge University Press (2010),
DOI: 10.1017/CBO9780511761362, ISBN 978-0-521-19249-1.

\bibitem{mayo2018}
D.~G.~Mayo,
\textit{Statistical Inference as Severe Testing: How to Get Beyond the Statistics Wars},
Cambridge University Press (2018),
DOI: 10.1017/9781107286184, ISBN 978-1-107-66464-7.

\bibitem{brilliant2026esp}
A.~M.~Brilliant,
``External Selection Under Correlated Error: Two Bounds and a Validation
Checklist for Generate-Then-Judge Workflows,''
TechRxiv preprint (2026),
DOI: 10.36227/techrxiv.176834656.66652387/v2.

\bibitem{osc2015}
Open Science Collaboration,
``Estimating the reproducibility of psychological science,''
Science \textbf{349}, aac4716 (2015),
DOI: 10.1126/science.aac4716.

\bibitem{nosek2018}
B.~A.~Nosek \textit{et al.},
``The preregistration revolution,''
Proc.\ Natl.\ Acad.\ Sci.\ USA \textbf{115}, 2600--2606 (2018),
DOI: 10.1073/pnas.1708274114.

\bibitem{mayospanos2006}
D.~G.~Mayo and A.~Spanos,
``Severe Testing as a Basic Concept in a Neyman--Pearson Philosophy of Induction,''
Br.\ J.\ Philos.\ Sci.\ \textbf{57}(2), 323--357 (2006),
DOI: 10.1093/bjps/axl003.

\bibitem{mayospanos2011}
D.~G.~Mayo and A.~Spanos,
``Error Statistics,''
in \textit{Philosophy of Statistics}, Handbook of Philosophy of Science, Vol.\ 7,
eds.\ P.~S.~Bandyopadhyay and M.~R.~Forster, Elsevier, pp.\ 153--198 (2011),
DOI: 10.1016/B978-0-444-51862-0.50005-8.

\bibitem{fletcher2020}
S.~C.~Fletcher,
``Of War or Peace? Essay Review of Statistical Inference as Severe Testing,''
Philos.\ Sci.\ \textbf{87}(4), 755--762 (2020),
DOI: 10.1086/709787.

\bibitem{cousins2020}
R.~D.~Cousins,
``Connections between statistical practice in elementary particle physics and the severity concept as discussed in Mayo's Statistical Inference as Severe Testing,''
arXiv:2002.09713 [physics.data-an] (2020).

\bibitem{lsnd2001}
A.~Aguilar \textit{et al.} (LSND Collaboration),
``Evidence for Neutrino Oscillations from the Observation of $\bar{\nu}_e$ Appearance in a $\bar{\nu}_\mu$ Beam,''
Phys.\ Rev.\ D \textbf{64}, 112007 (2001),
arXiv:hep-ex/0104049, DOI: 10.1103/PhysRevD.64.112007.

\bibitem{miniboone2018}
A.~A.~Aguilar-Arevalo \textit{et al.} (MiniBooNE Collaboration),
``Significant Excess of Electron-Like Events in the MiniBooNE Short-Baseline Neutrino Experiment,''
Phys.\ Rev.\ Lett.\ \textbf{121}, 221801 (2018),
arXiv:1805.12028 [hep-ex], DOI: 10.1103/PhysRevLett.121.221801.

\bibitem{icarus2013}
M.~Antonello \textit{et al.} (ICARUS Collaboration),
``Experimental search for the LSND anomaly with the ICARUS detector in the CNGS neutrino beam,''
Eur.\ Phys.\ J.\ C \textbf{73}, 2345 (2013),
arXiv:1209.0122 [hep-ex], DOI: 10.1140/epjc/s10052-013-2345-6.

\bibitem{diaz2020}
A.~Diaz, C.~A.~Arg\"uelles, G.~H.~Collin, J.~M.~Conrad, and M.~H.~Shaevitz,
``Where Are We With Light Sterile Neutrinos?''
Phys.\ Rep.\ \textbf{884}, 1--59 (2020),
arXiv:1906.00045 [hep-ex], DOI: 10.1016/j.physrep.2020.08.005.

\end{thebibliography}
\end{document}